\documentclass[preprint,proceedings]{rmaa}

\suppressfulladdresses 

\usepackage{paralist}

\usepackage{psfrag,color}

\SetYear{2007}
\SetConfTitle{The nuclear region, host galaxy and environment of 
active galaxies}

\title{Quasar Optical Variability and Black Hole Mass} 

\author{
  M. Wold,\altaffilmark{1} 
  M. S. Brotherton,\altaffilmark{2}
  and Z. Shang\altaffilmark{2}}

\altaffiltext{1}{Institute of Theoretical Astrophysics, University of Oslo,
  P.O. Box 1029 Blindern, N-0315 Oslo, Norway (wold@astro.uio.no).}

\altaffiltext{2}{Department of Physics and Astronomy, University of Wyoming, 
Laramie, WY 82072, USA.}

\shortauthor{Wold, Brotherton, \& Shang}
\shorttitle{RevMexAA(SC)}

\listofauthors{M. Wold, M. S. Brotherton, \& Z. Shang}
\indexauthor{Wold, M.}
\indexauthor{Brotherton, M.S.}
\indexauthor{Shang, Z.}

\abstract{In order to investigate the dependence of quasar optical-UV variability on fundamental 
physical parameters like black hole mass, we have matched quasars from the QUEST1 variability
survey with broad-lined objects from the SDSS. Black hole masses and bolometric luminosities
are estimated from Sloan spectra, and variability amplitudes from the QUEST1
light curves. Long-term variability amplitudes (rest-frame time scales 0.5--2 yrs) are found to 
correlate with black hole mass at the 99\% significance level or better. This means that 
quasars with larger black hole masses have larger percentage flux variations.
Partial rank correlation analysis shows that the correlation cannot explained by obvious selection effects inherent to flux-limited samples. We discuss whether the correlation is a manifestation of a relation between BH mass and accretion disk thermal time scales, or if it is due to changes in
the optical depth of the accretion disk with black hole mass. Perhaps the most likely explanation is that the more massive black holes are starving, and produce larger flux variations because they do not
have a steady inflow of gaseous fuel.}

\addkeyword{Quasars: variability}

\begin{document}
\maketitle

\section{Introduction}
\label{sec:intro}

Variability is a characteristic property of quasars. It is non-periodic in nature and shows erratic behaviour. Optical-UV variability (in unbeamed quasar continua) occurs on time scales of days to decades, with fluctuations typically a few tenths of a magnitude, increasing at longer time scales. 

Variability is recognized as an important diagnostic of the physical processes responsible for black 
hole activity. It is also important as the only method by which the smallest physical scales 
in AGN can be resolved.  The behaviour of quasar optical variability with luminosity and
redshift is well understood (see Vanden Berk et al.\ 2004 and references therein), but the real 
cause of the fluctuations remain opaque. In order to shed light on the origin of quasar variability,  relationships between variability and AGN fundamental parameters, such as black hole mass and Eddington ratio, may be investigated. 

Here we describe a study that correlates quasar long-term variability with BH mass.
Details are provided by Wold, Brotherton \& Shang (2007).

\section{Sample selection}
\label{sec:errors}

We form a sample of 104 quasars by matching broad emission line objects in the SDSS DR2 
(Abazajian et al.\ 2004) with sources in the 200k Light Curve Catalogue of the QUEST1 
variability survey (Rengstorf et al.\  2004). Redshifts were constrained to $z<0.75$ to keep the 
redshifted H$\beta$ line within the spectral coverage of the SDSS, and to
limit time dilation effects. 

\begin{figure}[!t]
  \includegraphics[width=6.5cm]{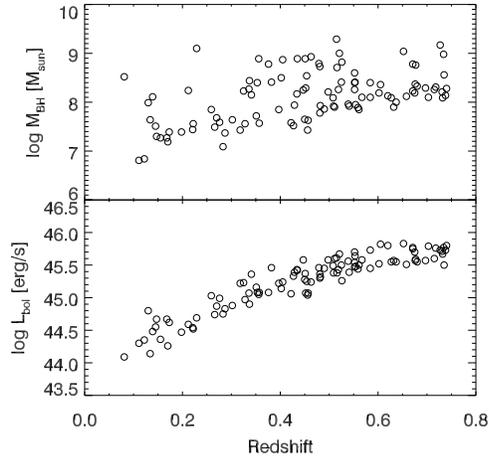}
  \caption{BH mass and bolometric luminosity as a function of redshift for the SDSS-QUEST1 sample.}
  \label{fig:fig1}
\end{figure}

From single epoch Sloan spectra we estimate virial BH masses using the H$\beta$  line width and the rest-frame quasar luminosity at 5100 \AA\@ following the scaling relationships of Vestergaard \& Peterson (2006). The bolometric luminosity of each quasar is evaluated as 
$L_{\rm bol} = 9\times\lambda L_{\lambda 5100} $ and the Eddington luminosity as 
$L_{\rm Edd} = 1.51\times10^{38}$ M$_{\rm BH}$/M$_{\odot}$ erg\,s$^{-1}$. Black hole mass and 
$L_{\rm bol}$ are plotted as a function of redshift in Figure~\ref{fig:fig1}.

\subsection{Variability measurements}

\begin{figure}[!t]
  \includegraphics[width=7cm]{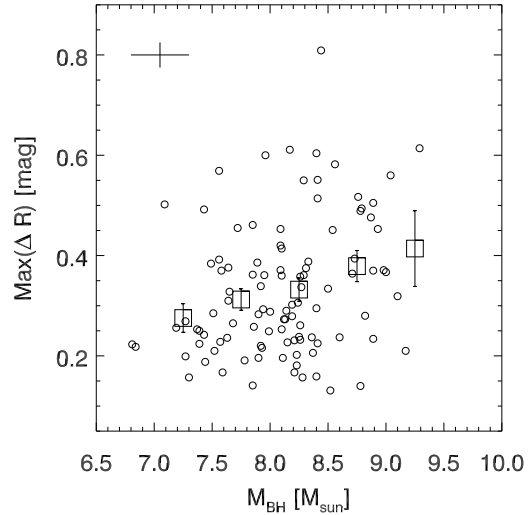}
  \caption{Variability amplitude as a function of BH mass. The mean 
  amplitude in five different BH mass bins is over-plotted as squares.}
      \label{fig:fig2}
\end{figure}

\begin{figure}[!t]
  \includegraphics[width=7cm]{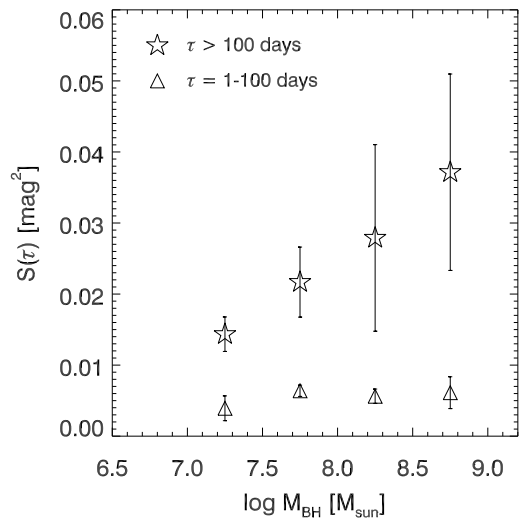}
  \caption{BH mass versus the structure function, $S(\tau)$, which measures
 how variability is distributed over rest-frame time scale $\tau$, here for time scales 1--100
  and $>100$ days (up to $\approx$2 years for the SDSS-QUEST1 sample). }
  \label{fig:fig3}
\end{figure}

Quasar variability is characterized by computing the distribution of all possible variability amplitudes (i.e.\ magnitude differences) on the quasar light curves, $\Delta m_{ij} = m_{i} - m_{j}$, 
where $i<j$. The variability of one quasar is described by the standard deviation, mean, 
median and/or maximum of its distribution of $\Delta m$ (e.g.\ Giveon et al. 1999). In Figure~\ref{fig:fig2}  we show the maximum variability amplitude for every quasar as a function of BH mass, and an 
increasing trend of variability with BH mass is seen.  Sources displaying the largest variability amplitudes have on average higher BH masses. The correlation is also illustrated in Figure~\ref{fig:fig3}, where it can be seen that most of the variability is detected at longer time scales. 

\section{Variability--BH mass correlation}

Pearson correlation analysis shows that the correlation between BH mass and variability is 
significant at the $\approx3\sigma$ level. The strongest correlation is measured between BH mass
and maximum variability (correlation coefficient 0.285 and a two-sided probability of arising by chance
of 0.3 \%). Other measures of variability  have qualitatively similar, but somewhat less
significant correlations with BH mass. 

There are no significant correlations detected between variability and Eddington ratio, 
$L_{\rm bol}/L_{\rm Edd}$, possibly because of the small sample size. The Eddington ratio is
clearly  an interesting parameter to correlate with variability as it measures whether variability
is related to how effectively the quasar converts accreted matter into luminosity 
(assuming a standard optically thick accretion disk as described by e.g.\ Shakura \& Sunyaev 1973).

\subsection{Selection effects}

The variability--BH mass correlation could be caused by more primary correlations with redshift and luminosity as these are known to correlate with variability (e.g.\ Vanden Berk et al.\ 2004). For instance, higher-$z$ quasars in the SDSS-QUEST1 sample are biased toward higher BH masses (Figure~\ref{fig:fig1}). A correlation between variability and BH mass could therefore be produced if underlying correlations existed between variability and $z$ (or between variability
and $L_{\rm bol}$). But Spearman's partial correlation analysis shows that this is not very likely. 
E.g.\ the correlation coefficient between variability and BH mass, at constant $z$, is 0.203 with a probability of 0.0035 of being caused by underlying trends with $z$. Similarly, 
Spearman's partial correlation coefficient between variability and BH mass, at constant $L_{\rm bol}$
is 0.254 with a probability of $2\times10^{-4}$ of occurring because
of underlying correlations with $L_{\rm bol}$. 

Can contamination by residual host galaxy light in the spectra artificially enhance a correlation
between variability and BH mass? Residual host galaxy light may dilute the variability and would be 
a potential problem only for the lower-luminosity quasars. However, only very few spectra
display clear host galaxy features, and these were treated extra carefully during the
spectral fitting. The variability--BH mass correlation is also present within a sub-sample of 
higher-luminosity quasars which are not expected to suffer from host contamination. 
Hence we conclude that although host contamination could artificially enhance the correlation,
it is a weak effect and unlikely to be important.

Finally, could the correlation be caused by time sampling effects? 
The best-sampled rest-frame time scale in the QUEST1 survey is 1--2 years and this is also where most of the variability is detected (Rengstorf et al.\ 2006; Wold et al.\ 2007). Figure~\ref{fig:fig3}
shows that the variability at shorter time scales (1--100 days) corresponds to the noise 
level in the data ($\approx 0.06$ mag), but that the variability at longer time scales is {\em intrinsic} to the quasars. 
Hence there is a bias toward detecting intrinsic quasar variability only at longer time scales. If a relation between BH mass and variability time scale exists, this may lead to a preference for detecting variability
only for quasars with higher BH masses.  And quasars with lower-mass BHs would have fluctuations
at or below the noise level, thus producing a variability--BH mass correlation.
There are certainly arguments that characteristic variability time scale (such as accretion disk
thermal time scale) may depend on BH mass
(e.g. Collier \& Peterson 2001), but this is still controversial and far from established. 

\section{Discussion}

We conclude that there is evidence for a correlation between the BH mass of a quasar and its 
optical long-term (0.5--2 yrs) variability properties. Quasars with more massive BHs tend to show 
larger variability amplitudes, i.e.\ quasars with larger BH masses have
larger percentage flux variations. Our favoured explanation is that the correlation is real and
not caused by selection effects. 
 
But what is its origin? Higher-mass BHs are more likely to be starving than those of lower mass.
This is because higher-mass BHs must have swallowed more of their surroundings in order to have grown to their larger masses. Hence the larger flux variations for the more massive BHs could indicate that the BH is running out of gaseous fuel. Because there is not a steady inflow of gaseous fuel from the surroundings, larger fluctuations in the continuum is seen. 

The correlation could also be explained in terms of optical depth of the accretion disk.
This can come about if the more massive BHs have accretion disks with larger optical depths
than those of lower mass. A large optical depth indicates radiatively more efficient accretion, 
hence any changes in the energy generation mechanism will be more easily transmitted through
an optically thick disk.

No significant correlations or anti-correlations could be detected between Eddington ratio and 
variability, probably because redshift and luminosity effects are difficult to disentangle in this sample.

A larger sample, or a sample spanning a longer time base line, is needed to 
confirm the general robustness of the variability--BH mass correlation.

\subsection{Acknowledgments}
MW thanks L. Binette and J.W. Sulentic for discussions.


\begin{thebibliography}

\bibitem{} Abazajian, K., et al. 2004, AJ, 128, 502

\bibitem{} Collier, S., Peterson, B.M. 2001, ApJ, 555, 775
\bibitem{} Giveon, U., Maoz, D., Kaspi, S., Netzer, H., Smith, P.S. 1999, MNRAS, 306, 637
\bibitem{} Rengstorf, A.W., et al., 2004, ApJ, 617, 184
\bibitem{} Rengstorf, A.W.,  Brunner, R.J., Wilhite, B.C., 2006, AJ, 131, 1923
\bibitem{} Shakura, N. I., Sunyaev, R. A. 1973, A\&A, 24, 337
\bibitem{} Vestergaard, M., Peterson, B.M. 2006, ApJ, 641, 689 
\bibitem{} Vanden Berk, D.E., et al. 2004, ApJ, 601, 692 
\bibitem{} Wold, M., Brotherton, M.S., \& Shang, Z. 2007, MNRAS, 375, 989

\end{thebibliography}
\end{document}